# Dominance of magnetoelastic coupling in multiferroic hexagonal YMnO$_3$

A. K. Singh and S. Patnaik

School of Physical Sciences,

Jawaharlal Nehru University New Delhi – 110067, India

S. D. Kaushik and V. Siruguri

UGC-DAE-Consortium for Scientific Research Mumbai Centre,

R5 Shed, Bhabha Atomic Research Centre, Mumbai – 400085, India

Corresponding Author

spatnaik@mail.jnu.ac.in

**Abstract**

Hexagonal $YMnO_3$ with space group *P6₃cm* is one of the rare examples of multiferroic materials that exhibit co-dependence of a robust ferroelectric phase along with antiferromagnetic ordering. This is primarily ascribed to tilting of the $MnO_5$ polyhedra driven by magnetoelectric coupling. We report on the effect of magnetic field at the atomic level in $YMnO_3$ by utilizing magnetic field and temperature dependent neutron diffraction measurements. We show that near the Néel temperature the lattice parameters, effective magnetic moment per site, Mn-O bond lengths and O-Mn-O bond angles show large variations and these are further modified in the presence of magnetic field. We observe distinct changes with the application of magnetic field in the paramagnetic state both in atomic positions and in the bulk dielectric constant. Our results provide unambiguous confirmation of the role played by exchange-striction over and above the much studied magnetoelectric coupling in this frustrated antiferromagnet.

**Introduction**

Several recent discoveries involving magnetism driven ferroelectricity and new insight into the intertwined states of electric, magnetic and elastic order parameters have redefined the broad scope of research in multiferroic materials. [1-7] Such multifunctional materials are of great technological significance; potentially it could lead to eight-state memory devices, [8] gate ferroelectrics based field-effect transistors, [9] and electric field controlled magnetic resonance devices. [10-14] Furthermore, because of the associated rich magnetic-electric-elastic phase diagram, new functionalities and their theoretical foundation are under intense investigation. [15, 17] In this context, the rare earth manganites ($RMnO_3$) have attracted considerable attention. Both orthorhombic (for large ionic radius R = La, Pr, Nd, Sm, Eu, Gd, Dy and Tb; space group *Pbnm*) [18, 19], as well as hexagonal (for smaller ionic radius R = Ho, Er, Tm, Yb, Lu, Sc and Y; space group *P6$_3$cm*) [20] phases of $RMnO_3$ exhibit multiferrocity but the microscopic origins are different. In orthorhombic case, the magnetic frustration leads to spin lattice coupling which is induced by inverse Dzyaloshinski-Moriya (DM) interaction, [5] whereas in the hexagonal case, tilting of $MnO_5$ bipyramids and buckling of rare earth layers are responsible for multiferroicity. [21] In particular, the hexagonal phase of $YMnO_3$ has been treated as the prototype for a microscopic understanding of the magnetoelectric (ME) effect. [22-26] It shows antiferromagnetic ordering below $T_N \sim 65$ K with moments aligned in the ab plane and a ferroelectric transition at $T_C \sim 900$ K. [27, 28] Detailed zero field neutron scattering suggests that all structural parameters change at $T_N$ implying strong spin-lattice coupling and consequent magnetoelectric coupling. [22, 26] It was shown that electric polarization across the antiferromagnetic transition showed the same temperature dependence as the effective ordered moment μ assigned to $Mn^{3+}$ state. But very recent studies have indicated giant magnetoelastic coupling in $YMnO_3$, almost two orders of magnitude larger than any other magnetic material and thus leading to large scale displacement of all the atoms in the unit cell. [26] This sets a question as to whether the magnetic field dependence of dielectric constant in $YMnO_3$ is caused by magnetoelectric coupling, that necessarily requires an ordered magnetic state, or due to exchange-striction driven magneto-elastic coupling. Using magnetic field dependent neutron scattering we show that it is magnetoelasticity that dominates over very weak magnetoelctric coupling in hexagonal $YMnO_3$.

The structure of hexagonal $YMnO_3$, as depicted in Fig. 1, consists of close-packed layers of bipyramidal $MnO_5$ where each Mn ion is surrounded by three equatorial (in-plane) and two apical $O^{2-}$ ions. Since $Y^{3+}$ ($4p^6$) is non magnetic, the magnetic property in $YMnO_3$ is controlled by the $Mn^{3+}$ ($3d^4$) with spin S=2. The $MnO_5$ bipyramids are corner linked to form a triangular lattice in the *ab*-plane and are separated from one another along *c*-axis by rare earth layers and hence theories relating to 2D magnetism have been invoked. [29, 30] Due to antiferromagnetic interaction between the spins of the close-packed $Mn^{3+}$ ions, located in the z=0 and z=1/2 planes, the moments get aligned into a $120^\circ$ structure in the *ab* plane. [25, 31] Consequent upon the 2D magnetism and triangular lattice, the magnetic structure in $YMnO_3$ is frustrated. The ferroelectric polarization on the other hand appears along the *c*-axis perpendicular to the Mn-O plane [32] and it is understood that tilting of $MnO_5$ trigonal bipyramids and buckling of Y layers are the source of spontaneous electric polarization in the absence of external electric field. [21-27]

**Experiments**

The polycrystalline samples investigated were prepared by conventional solid-state sintering process at ambient pressure. High purity powders of cation oxides $Y_2O_3$ (99.99%) and $MnO_2$ (99.99%) were thoroughly mixed in a ratio of 1:2 to achieve the stoichiometry of $YMnO_3$. Then it was compacted and calcinated at 1100 $^o$C for 13 hours. To ensure homogeneity and density of pellets, the sintered pellets were ground, compacted and reheated twice at 1300 $^o$C for 12 hours. Room temperature powder x-ray diffraction (XRD) studies of the samples were performed using a Bruker D8 X-ray diffractometer with Cu K$\alpha$ radiation ($\lambda$=1.5406 Å). The neutron diffraction measurements were carried out on powder samples using the multi-PSD Focusing Crystal Diffractometer (FCD) set up by UGC-DAE Consortium for Scientific Research Mumbai Centre at the National Facility for Neutron Beam Research (NFNBR), Dhruva reactor, Mumbai (India) at a wavelength of 1.48 Å. Neutron diffraction (ND) patterns were recorded at 10, 30, 60, 100 and 300 K. ND patterns were also recorded at 30, 60 and 100 K in an external magnetic field of 5 T. The samples were placed in vanadium cans that were directly exposed to neutron beam for 300 K data. For low temperature and in-field neutron data, vanadium cans filled with the powder samples were loaded in a Cryogen-Free Magnet System (CFM). The data were analysed using Rietveld method and the refinement of both crystal and magnetic structures was carried out using the FullProf program. [33] Temperature and frequency dependent dielectric measurements on the pellets were performed using a QUADTECH 1920 precision LCR meter.[35] Resistivity of the insulating samples was measured using a constant voltage source in a two-probe configuration. Magnetization and specific heat data were recorded in Quantum Design MPMS squid and PPMS respectively.

**Results & discussion**

All the peaks in room temperature XRD pattern for $YMnO_3$ were satisfactorily indexed in hexagonal crystal structure with space group *P6₃cm*. Energy dispersive X-ray spectroscopy (EDAX) measurement confirmed the stoichiometry and within the experimental error, the molar ratio of Y: Mn: O was found to be 1: 1: 3. Fig. 2 shows the reciprocal magnetic susceptibility ($1/\chi$) as a function of temperature for $YMnO_3$ which indicates $T_N \sim 62$ K. The extrapolated Curie-Wiess temperature ($\Theta_{CW}$) is found to be $\sim -330$ K and the ratio of $\Theta_{CW}$ and $T_N$, that relates to the degree of frustration, is estimated $\sim 5.3$. The zero field specific heat data plotted in the inset of Fig. 2, shows sharp deviation at $T_N$. After subtracting the phonon contribution, the zero field magnetic entropy is estimated to be $\sim 11.45$ J/mol-K up to 90 K. [30]

Fig. 3a shows Rietveld refinement of room temperature ND data that confirm the hexagonal structure of $YMnO_3$ in space group *P6₃cm*. Lattice parameters *a*= 6.149 Å and *c*= 11.363 Å are in excellent agreement with published data. [25-27] ND pattern does not show the presence of any secondary phase thus confirming the phase purity. Structural parameters after the Rietveld refinement of neutron diffraction data for $YMnO_3$ at room temperature are shown in the Table 1. In the table, *Biso* is the isotropic Debye-Waller factor (one of the refinement parameters). Zero field ND data were taken at 10, 30, 60, 100 and 300 K and there is no evidence of structural change from hexagonal structure. We observe that both the lattice constants *a* and *c* show significant changes near Néel temperature. Lattice constant *a* and the unit cell volume show positive thermal expansion whereas lattice constant *c* shows negative thermal expansion. Previous studies have reported this anomalous behavior of negative thermal expansion of **c**-axis to persist above the ferroelectric transition. [27]

Magnetic field dependent ND has been carried out at 30 K corresponding to a bulk antiferromagnetic state, at 60 K to a crossover state and at 100 K to a paramagnetic state. In Fig. 3b we show the refined ND pattern at 60 K for at 0 and 5 T magnetic field. ND data were taken in the warming cycle. After collecting the data at the particular temperature, field was reduced to zero and sample temperature was raised well beyond the Néel temperature to erase the possibility of field induced spin arrest. We observe a clear signature of enhancement in (002) Bragg peak in ND data at 60 K (not observed at lower temperatures) in the presence of 5 T magnetic field. The (002) peak is nuclear in nature and does not carry any magnetic contribution in the AF-ordered structure of $YMnO_3$. Hence, the origin of this enhancement is attributed to preferred orientation of grains possibly brought about by the presence of weak field-induced ferromagnetic correlations. Peaks like (006) and (00 10) also show increase in intensity. We invoked the preferred orientation parameter in the Reitveld refinement, which resulted in an excellent fit to the data (Fig. 3b). The gap at around 60 degrees is due to interference peaks from the magnet shroud.

The temperature and field variation of all the parameters studied are summarized in Fig. 4. It is known that the refinement of magnetic structures in powder samples is quite complicated in the ordered state owing to the fact that the magnetic grains may have different magnetic structures depending on the field orientation with respect to the grains. Hence, in the absence of any previous data in presence of external magnetic field either in powder or single crystal, the present results on polycrystalline samples in the ordered state (<60 K) represent an average picture of the spin structure. ND patterns taken at 10 K and 30 K show the expected magnetic intensities, in particular the peak (101) which is forbidden by the space group $P6_3cm$. Appearance of this peak is a signature of the antiferromagnetic ordering below $T_N$ characterized by the propagation vector $k$=0. For the magnetic refinement of the ND data, the structure described by the basis vectors of the irreducible representation $\Gamma_1$ was used. Since no new peaks were observed in the presence of external magnetic field, it is deduced that the basic magnetic structure is not changed and hence, the same irreducible representation was used to fit the in-field ND data.

We observe that at 30 K, for 5 T magnetic field, lattice constant $c$ decreases compared to zero field value. This value suppression becomes larger at 100 K (Fig. 4b). On the other hand, the lattice parameter $a$ decreases with respect to zero field value at 30 K but near $T_N$ the value increases significantly (Fig. 4a). Thus in the presence of magnetic field at temperature above and below $T_N$, both the lattice parameters get altered as compared to zero field value. We note that there is no orbital degree of freedom as the 4 d electrons of $Mn^{3+}$ occupy the low energy doublets (xz, yz and xy, $x^2$-$y^2$) and hence the changes in lattice parameters are not due to Jahn–Teller distortion. [35] These large changes in the paramagnetic state with the application of field that can only be attributed to strong magneto-elastic coupling. We also note that the zero field effective moment obtained from ND data at 30 K is estimated to be 2.9 $\mu_B$, much smaller than ionic value of 4 $\mu_B$. This reduction in Mn magnetic moment is because of 2D geometrical frustration in the Mn lattice. The Mn magnetic moment reduces as temperature is raised from 10 K to 60 K and as shown in Fig. 4d, in the presence of 5 T magnetic field, effective Mn moment decreases further. The ordered Mn moment decreases by ~ 40 % near Néel temperature as compared to zero field value. This indicates increased frustration amongst Mn spins with the application of magnetic field.

The field dependent variation of lattice parameters as a function of temperature gives indication that the tilting of $MnO_5$ polyhedra can be tuned by a relatively small field of 5 T. To investigate this possibility, we discuss next how the magnetic field affects the Mn-O bond distances and O-Mn-O bond angles both in the antiferromagnetic and paramagnetic states and correlate this to the magnetic field dependent dielectric constant.

In the paramagnetic state, the Mn atom is located at x ≅ 1/3 in the geometry of a two dimensional triangular lattice. With the onset of ordering at $T_N$, it is reported that the position of Mn atom shifts by 3.3 % from x ≅ 1/3 value. [22,23,24] Similar trend is observed in our analysis on ND data in absence of magnetic field. The variation in the Mn atomic position is ~ 3% from 300 K to 10 K. We further observe that in zero field, all Mn-O bond distances show significant variation at $T_N$. As the temperature decreases below Néel temperature, the Mn and apical oxygen ($O_{ap}$) bond lengths increase whereas Mn and equatorial oxygen ($O_{eq}$) bond lengths decrease. [26, 28]. With the application of 5 T magnetic field, all the Mn-O bond lengths get further modified. The magnetic field affects all the Mn-O bonds whether it is $O_{eq}$ (O(3) and O(4)) or $O_{ap}$ (O(1) and O(2)). These changes seen in all four Mn-O bond distances as a function of temperature and magnetic field are shown in Fig. 4e-h. Clearly at 30 K, magnetic field decreases the Mn-$O_{ap}$ bond length and this change is more pronounced above $T_N$. On the other hand, at 30 K, the magnetic field decreases the Mn-$O_{eq}$ bond length but near $T_N$, it increases considerably as compared to zero field value. These variations are also seen when O-Mn-O bond angles are calculated from Rietveld refinement of powder ND data (shown in Fig. 4i-m). We conclude that application of magnetic field both in the paramagnetic and antiferromagnetic states, lead to decrease in distance between Mn and $O_{ap}$ (O(1) and O(2)) where as $O_{eq}$ (O(3) and O(4)) go farther from Mn. This leads to reduction in the tilting of $MnO_5$ polyhedra. Consequently, it is expected that polarization in presence of magnetic field will decrease as compared to zero field value both below and above $T_N$. Refined *x* positions of Mn atom as a function of temperature in zero magnetic field is shown in Fig. 4n.  The position of Mn does not vary significantly in the neutron data recorded in the presence of magnetic field.

Fig. 5 shows the temperature dependent dielectric constant ($\varepsilon$) of a sintered pellet of $YMnO_3$ across the antiferromagnetic transition at a frequency of 1 kHz and at $\mu_0H = 0$, 3, and 5 T.  To rule out the possibility of leakage current dominating the measurement, in the top inset we show the resistivity as a function of temperature both at zero field and at $\mu_0H = 4$ T.  There is no trace of magnetoresistance, so one can safely assign the changes in the dielectric constant with the magnetic field to be of capacitive origin. [7, 36] With zero external field, a broad shoulder appears in the vicinity of ~ 62 K which is a signature of a weak first order phase transition (antiferromagnetic ordering). [7, 37]  With the application of 3 T magnetic field, the dielectric constant decreases and this is more when 5 T magnetic field is applied.  The estimated magnetocapacitance (($\varepsilon$ (T, H) − $\varepsilon$ (T, 0)/ $\varepsilon$ (T, 0) × 100) at T = 65 K are 0.33 and 0.5 for magnetic fields of 3 T and 5 T respectively.  Clearly the magnetic field suppresses the tilting of $MnO_5$ polyhedra in the antiferromagnetic state as indicated in our field dependent neutron scattering data. Further, the observed magnetocapcitance for 5 T magnetic field at 30, 60 and 90 K are of comparable magnitude, 0.552, 0.528 and 0.424 respectively.   For pure magneto-electric coupling (change in polarization due to change in magnetic state of $Mn^{3+}$), induced polarization P goes as ~ $\mu_{eff}^2$ where $\mu_{eff}$ is the magnetic moment[22, 26].   Since effective moment decreased by 40 % at 5 T near $T_N$, the corresponding extrapolated $\Delta$P would be substantially larger than what is observed experimentally near $T_N$. Most importantly, we observe that the magnitude of dielectric constant decreases in the paramagnetic state where there is no ordered magnetic state and the magnetocapacitance values are similar both below and above $T_N$. It again reinforces the interpretation that dominant changes in the presence of magnetic field, both in the antiferromagnetic and in the paramagnetic states, have the origin in magnetoelasticity. In essence, the spin exchange interactions in $YMnO_3$ are frustrated and the spins in the system can undergo a transition into an ordered AF state by releasing frustration through the lattice displacements. This provides a suitable mechanism for the magnetodielectric effect in a frustrated spin system. [38] We note that the multiferrocity reported in the incommensurate spin frustrated system such as orthorhombic $TbMnO_3$ and $YMnO_3$ are in conformity with this picture.

**Conclusions**

In conclusion, high quality polycrystalline samples of hexagonal $YMnO_3$ have been extensively characterized using magnetization, specific heat, dielectric constant, and magnetic field dependent neutron diffraction. At the microscopic level, the changes in lattice constant, Mn moment, Mn-O bond length and O-Mn-O bond angle are quantitatively estimated both above and below the Néel temperature. A relatively small field of 5 Tesla is found sufficient to usher in significant changes in atomic positions and in dielectric constant both in the antiferromagnetic as well as the paramagnetic state. This is reflective of the dominant role played by magnetoelastic coupling vis á vis weak magnetoelectric coupling in this extensively studied multiferroic compound.

**Acknowledgement**

We thank the Department of Science of Technology, Government of India, and the Consortium for Scientific Research, India for providing support through a Collaborative Research Scheme for accessing their neutron diffraction facility at NFNBR, BARC, Mumbai. SBP would thank Deepak Kumar for very useful discussions. We thank Dr. V. Ganesan, UGC-DAE Consortium for Scientific Research, Indore, India for the specific heat measurement. DST, Government of India is acknowledged for funding the SQUID at IIT Delhi. A.K.S. acknowledges CSIR, India for financial assistance.

**References:**

1. Nicola A. Spaldin and Manfred Fiebig, Science **309**, 391 (2005).

2. Daniel Khomskii, Physics **2**, 20 (2009).

3. D. Chiba, M. Sawicki, Y. Nishitani, Y. Nakatani, F. Matsukura & H. Ohno, Nature, **455**, 515 (2008).

4. T. Kimura, T. Goto, H. Shintani, K. Ishizaka, T. Arima & Y. Tokura, Nature **426**, 55 (2003).

5. Sang - Wook Cheong and Maxim Mostovoy, Nat. Mater. **6**, 13 (2007).

6. Jeroen van den Brink and Daniel I khomskii, J. Phys.: Condens. Matter **20**, 434217 (2008).

7. A. K. Singh, Michael Snure, Ashutosh Tiwari and S. Patnaik, J. Appl. Phys. **106**, 014109 (2009).

8. F. Yang, M. H. Tang, Z. Ye, Y. C. Zhou, X. J. Zheng, J. X. Tang, J. J. Zhang, and J. He, J. Appl. Phys. **102**, 044504 (2007).

9. M. Fiebig, T. Lottermoser, D. Fröhlich, A. V. Goltsev, & R. V. Pisarev, Nature **419**, 818 (2002).

10. J. Seidel, L.W. Martin, Q. He, Q. Zhan, Y. H. Chu, A. Rother, M. E. Hawkridge, P. Maksymovych, P. Yu, M. Gajek, N. Balke, S. V. Kalinin, S. Gemming, F.Wang, G. Catalan, J. F. Scott, N. A. Spaldin, J. Orenstein and R. Ramesh, Nat. Mater. **8**, 229 (2009).

11. F. Moussa, M. Hennion, J. Rodriguez-Carvajal, and H. Moudden, Phys. Rev. B **54**, 15149 (1996).

12. Yoshinori Tokura, Science **312**, 1481 (2006).

13. Marian Vopsaroiu, John Blackburn, A. Muniz-Piniella, and Markys G. Cain, J. Appl. Phys. **103**, 07F506 (2008).

14. Junyi Zhai, Zengping Xing, Shuxiang Dong, Jiefang Li, and D. Viehland, Appl. Phys. Lett. **88**, 062510 (2006).

15. Special issue, J. Phys. Cond. Mat. 20, 434201–434220 (2008).

16. T. Kimura, T. Goto, H. Shintani, K. Ishizaka, T. Arima & Y. Tokura, Nature **426**, 55 (2003).

17. R. Ramesh and Nicola A. Spaldin, Nat. Mater. **6**, 21 (2007).

18. Kunihiko Yamauchi, Frank Freimuth, Stefan Blügel, and Silvia Picozzi, Phys. Rev. B **78**, 014403 (2008).

19. B. Lorenz, Y. Q. Wang, Y. Y. Sun, and C. W. Chu, Phys. Rev. B **70**, 212412 (2004).

20. H.L. Yakel, W.C. Koehler, E.F. Bertant, E.F. Forrat, Acta Crystallogr. **16,** 957 (1963).

21. Bas B. Van Aken, Thomas T.M. Palstra, Alessio Filippetti and Nicola A. Spaldin, Nat. Mater. **3**, 164 (2004).

22. Seongsu Lee, A. Pirogov, Misun Kang, Kwang-Hyun Jang, M. Yonemura, T. Kamiyama, S.-W. Cheong, F. Gozzo, Namsoo Shin, H. Kimura, Y. Noda & J.-G. Park, Nat. **451**, 805 (2008).

23. X. Fabrèges, S. Petit, I. Mirebeau, S. Pailhès, L. Pinsard, A. Forget, M. T. Fernandez-Diaz, and F. Porcher, Phys. Rev Lett. 103, 067204 (2009).

24. I. Munawar, and S. H. Curnoe, J. Phys. Condens. Matter 18, 9575 (2006).

25. A. Muñoz, J. A. Alonso, M. J. Martinez-Lope, M. T. Casais, J. L. Martinez, and M. T. Fernandez-Diaz, Phys. Rev. B **62**, 9498 (2000).

26. Seongsu Lee, A. Pirogov, Jung Hoon Han, J.-G. Park, A. Hoshikawa, and T. Kamiyama, Phys. Rev. B **71**, 180413 (2005).

27. T. Katsufuji, M. Masaki, A. Machida, M. Moritomo, K. Kato, E. Nishibori, M. Takata, M. Sakata, K. Ohoyama, K. Kitazawa, and H. Takagi, Phys. Rev. B **66**, 134434 (2002).

28. M. Fiebig, D. Fröhlich, K. Kohn, St. Leute, Th. Lottermoser, V. V. Pavlov, and R.V. Pisarev, Phys. Rev. Lett. **84**, 5620 (2000).

29. S. Petit, F. Moussa, M. Hennion, S. Pailhés, L. Pinsard-Gaudart, and A. Ivanov, Phys. Rev. Lett. **99**, 266604 (2007).

30. Junghwan Park, Misun Kang, Jiyeon Kim, Seongsu Lee, Kwang-Hyun Jang, A. Pirogov, J.-G. Park, Changhee Lee, S.-H. Park and Hyoung Chan Kim, Phys. Rev. B **79**, 064417 (2009).

31. Y. Aikawa, T. Katsufuji, T. Arima, and K. Kato, Phys. Rev. B **71**, 184418 (2005).

32. S. Pailhès, X. Fabrèges, L. P. Régnault, L. Pinsard-Godart, I. Mirebeau, F. Moussa, M. Hennion, and S. Petit, Phys. Rev. B **79**, 134409 (2009).

33. J. Rodriguez-Carvajal, Physica B **192**, 55 (1993).

34. S. D. Kaushik, Anil K. Singh, D. Srikala and S. Patnaik, Indian J Pure  and Appl. Phys. **46**, 334 (2008)

35. T. Mizokawa, D. I. Khomskii, and G. A. Sawatzky, Phys. Rev. B **60**, 7309 (1999).

36. G. Catalan, Appl. Phys. Lett. **88**, 102902 (2006).

37. Z. J. Huang, Y. Cao, Y. Y. Sun, Y. Y. Xue, and C. W. Chu, Phys. Rev. B **56**, 2623 (1997).

38. A. K. Singh, S. D. Kaushik, B. Kumar, P. K. Mishra, A. Venimadhav, V. Siruguri, and S. Patnaik, Appl. Phys. Lett. 92, 132910 (2008).

**Figure Caption:**

**Fig. 1.** (Color online) **(a)** Schematic representation of the crystal structure of hexagonal $YMnO_3$. Here apical oxygen ($O_{ap}$) and equatorial oxygen ($O_{eq}$) atoms are represented with different symbols. Mn atoms are situated at z=0, z=1/2 and z=1 plane and each Mn is surrounded by five oxygen atoms forming $MnO_5$ trigonal bipyramidal structure. O1, O2 represent apical oxygen ($O_{ap}$) atoms and O3, O4 represent equatorial oxygen ($O_{eq}$) atoms.

**Fig. 2.** (Color online) Temperature variation of the inverse magnetic susceptibility measured in zero field-cooled (ZFC) conditions under $\mu_0 H = 0.5$ T. Inset shows the zero field specific heat. Anomaly in both $\chi$ vs T as well as specific heat at ~62 K is a signature of antiferromagnetic ordering.

**Fig. 3.** (Color online) **(a)** Neutron-diffraction data (symbols) taken at 300 K for $YMnO_3$. The solid line represents the calculated pattern with hexagonal ($P6_3cm$) symmetry. **(b)** Neutron powder diffraction patterns of $YMnO_3$ at 60 K (near $T_N$) recorded at $\mu_0 H = 0$ T and 5 T. The blue lines in the figure indicate the difference between the observed and calculated diffraction patterns. The bars indicate the positions of the nuclear Bragg peaks in (a), and at 60 K (b), the upper and lower sets of bars correspond to nuclear and magnetic phases, respectively. R factors for fitting at 60 K are $\chi^2(0T)=2.23$, Rp $(0T)=2.3\%$, Rwp(0T)=3.12\%, $R_{exp}=2.09\%$ and $\chi^2(5T)=4.25$, Rp(5T)=3.32\%, Rwp(5T)=4.29\%, $R_{exp}=2.08\%$.

**Fig. 4.** (Color online) Temperature and field dependence of **(a)** lattice constant *a*, **(b)** lattice constant *c*, **(c)** unit cell volume, **(d)** ordered magnetic moment per Mn atom, **(e-h)** Mn-O bond lengths, and **(i-m)** O-Mn-O bond angles. The O-Mn-O bond angles

are measured in degrees. **(n)** *x* position of Mn atom as a function of temperature. Values are obtained from refinement of neutron powder diffraction data at 30, 60, and 100 K in zero field (■) and 5 T (•) magnetic field. O1, O2 represent apical oxygen atoms and O3, O4 represent equatorial oxygen atoms.

**Fig.5.** (Color online) Temperature dependence of dielectric constant measured at different magnetic field $\mu_0 H = 0$ T (■), 3 T (✳) and 5 T (▲) taken at 1 kHz frequency in the temperature range 25-105 K. Inset (a) shows temperature dependent dielectric constant in the temperature range 10-265 K. Change in slope at T ~ 62 K reflects the antiferromagnetic ordering. Inset (b) shows temperature dependence of resistivity ($\rho$) at $\mu_0 H = 0$ T (■) and 4 T (•).

**Table caption**

**Table 1.** Structural parameters after the Rietveld refinement of neutron diffraction pattern of $YMnO_3$ at room temperature. Atomic positions of Y1 and O3 at 2a (0,0,z); Y2 and O4 at 4b (1/3, 2/3,z); Mn, O1, and O2 at 6c (x,0,z) (for Mn, z = 0).

**Table -1**

| Parameters | | |
|---|---|---|
| a (Å) | | 6.1490 |
| c (Å) | | 11.3627 |
| V (Å$^3$) | | 372.073 |
| Y1 | z | 0.27206 |
| | Biso | 0.422 |
| Y2 | z | 0.22900 |
| | Biso | 0.422 |
| Mn | x | 0.32472 |
| | Biso | 0.498 |
| O1 | x | 0.30952 |
| | Biso | 0.450 |
| | z | 0.15127 |
| | Biso | 0.450 |
| O2 | x | 0.64491 |
| | Biso | 0.445 |
| | z | 0.32599 |
| | Biso | 0.445 |
| O3 | z | 0.47814 |
| | Biso | 0.795 |
| O4 | z | 0.01227 |
| | Biso | 0.795 |
| Discrepancy factors | | |
| $\chi^2$ | | 3.93 |
| $R_p$ (%) | | 2.17 |
| $R_{wp}$ (%) | | 2.89 |
| $R_{exp}$(%) | | 1.23 |

**Fig. 1**

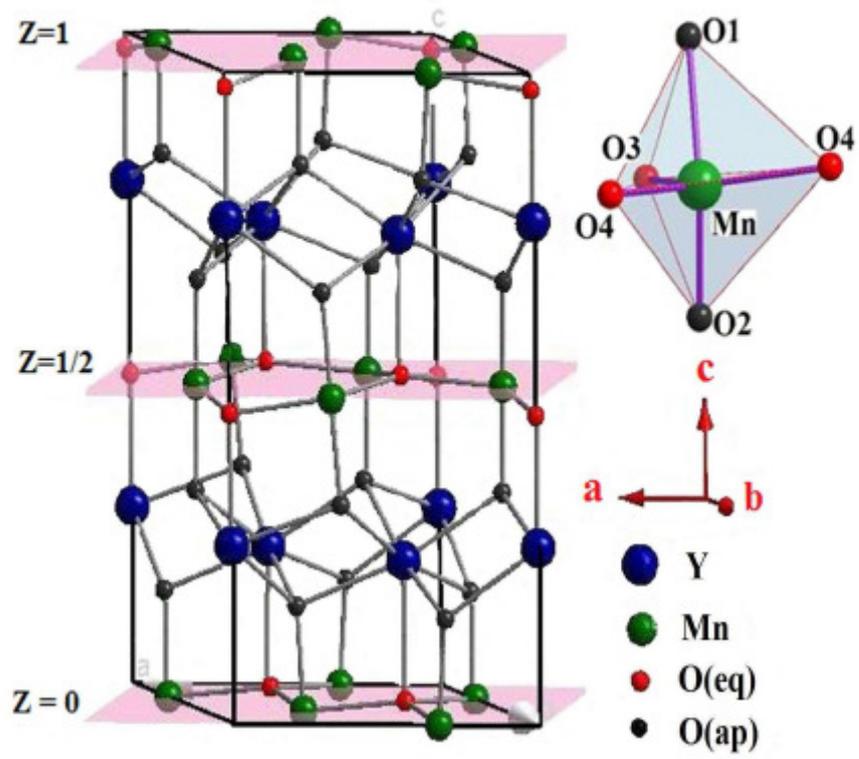

**Fig. 2**

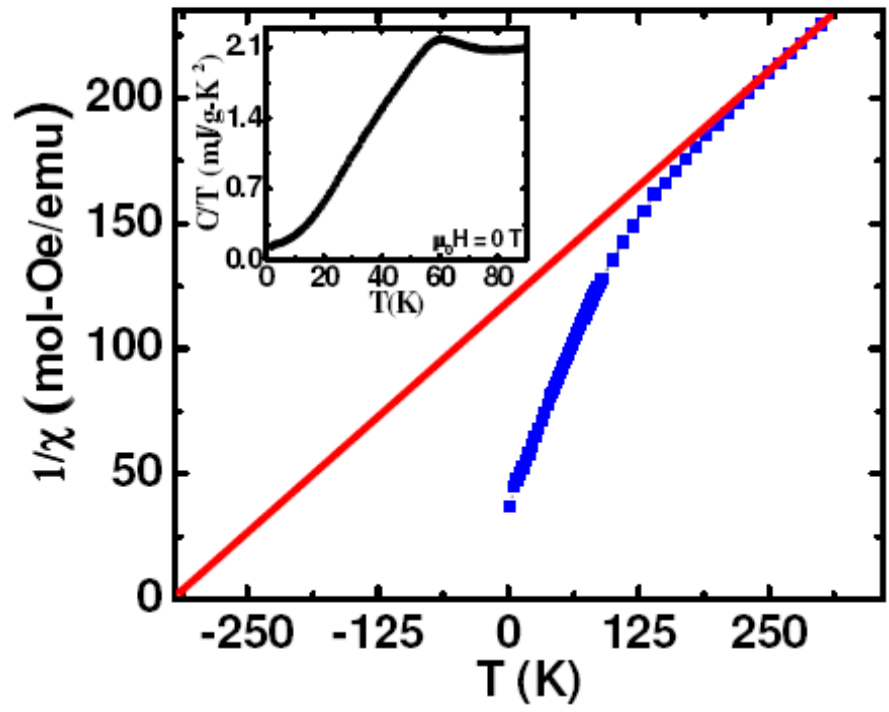

**Fig. 3a**

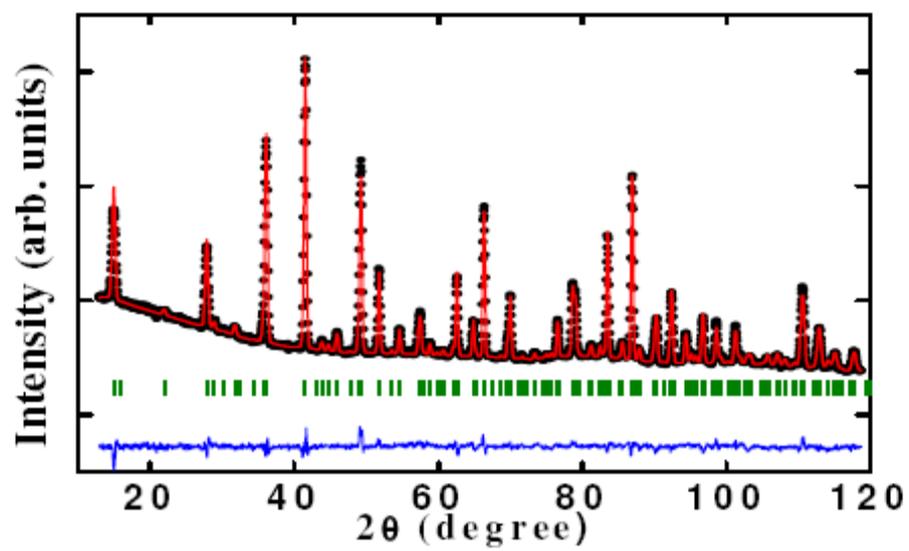

**Fig. 3b**

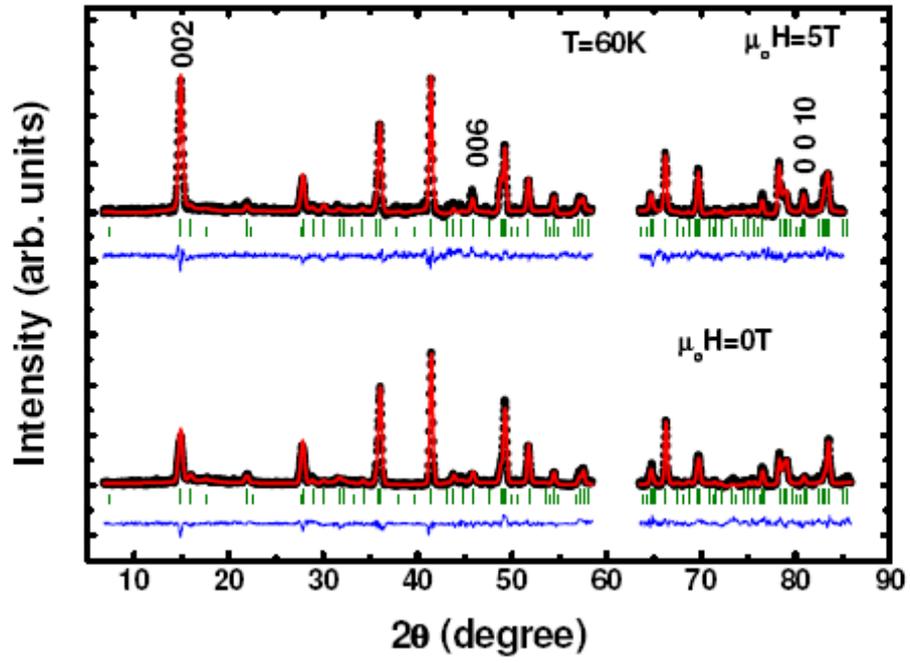

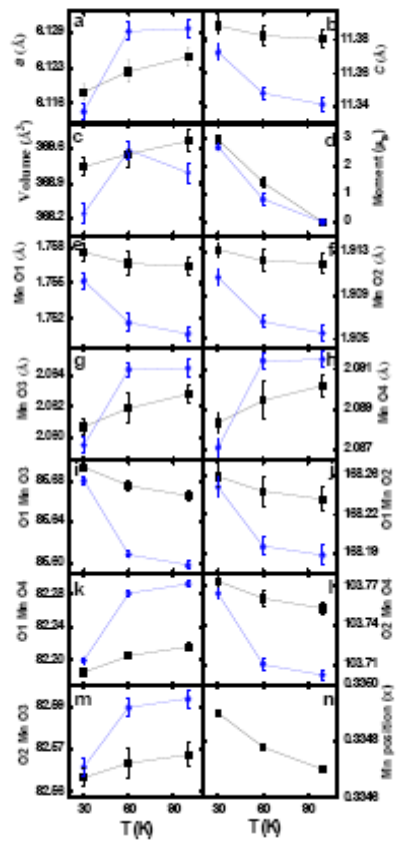

**Fig. 5**

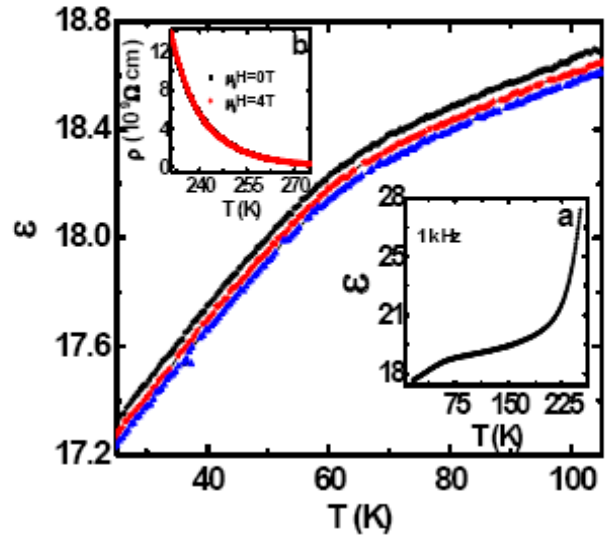